\documentclass[twocolumn,pra,aps,showpacs]{revtex4}
\usepackage{amsmath}
\usepackage{color}
\usepackage{graphicx}
\usepackage{CJK}

\begin{document}
\begin{CJK*}{GBK}{song}

\title{ Optical properties of a waveguide-mediated chain of randomly positioned atoms }   
\author{Guo-Zhu Song,$^{1,}$\footnote{Corresponding author: songguozhu@tjnu.edu.cn } Jin-Liang Guo,$^{1}$ Wei Nie,$^{2}$ Leong-Chuan Kwek,$^{3,4,5,6}$ and Gui-Lu Long$^{7,8,9,10}$ }

\address{$^{1}$College of Physics and Materials Science, Tianjin Normal University, Tianjin 300387, China\\
$^{2}$Theoretical Quantum Physics Laboratory, RIKEN Cluster for Pioneering Research, Wako-shi, Saitama 351-0198, Japan\\
$^{3}$Centre for Quantum Technologies, National University of Singapore, 3 Science Drive 2, Singapore 117543\\
$^{4}$Institute of Advanced Studies, Nanyang Technological University, Singapore 639673\\
$^{5}$National Institute of Education, Nanyang Technological University, Singapore 637616\\
$^{6}$MajuLab, CNRS-UNS-NUS-NTU International Joint Research Unit, UMI 3654, Singapore\\
$^{7}$State Key Laboratory of Low-Dimensional Quantum Physics and Department of Physics, Tsinghua University, Beijing 100084, China\\
$^{8}$Beijing National Research Center for Information Science and Technology, Beijing 100084, China\\
$^{9}$Frontier Science Center for Quantum Information, Beijing 100084, China\\
$^{10}$Beijing Academy of Quantum Information Sciences, Beijing 100193, China  }
\date{\today }

\begin{abstract}

We theoretically study the optical properties of an ensemble of two-level atoms coupled to a one-dimensional waveguide. In our model, the atoms are randomly located in the lattice sites along the one-dimensional waveguide.
The results reveal that the optical transport properties of the atomic ensemble are influenced by the lattice constant and the filling factor of the lattice sites. We also focus on the atomic mirror configuration and quantify the effect of the inhomogeneous broadening in atomic resonant transition on the scattering spectrum. Furthermore, we find that initial bunching and persistent quantum beats appear in photon-photon correlation function of the transmitted field, which are significantly changed by filling factor of the lattice sites. With great progress to interface quantum emitters with nanophotonics, our results should be experimentally realizable in the near future.

\end{abstract}
\pacs{03.67.Lx, 03.67.Pp, 42.50.Ex, 42.50.Pq}

\maketitle

\section{Introduction}

In the past decades, waveguide quantum electrodynamics (QED) has raised great interest owing to its promising applications in
quantum devices and quantum information technologies \cite{ShenOL2005,Shen2007PRL,Zhou2013prl,PLodahl2015rmp,Liao2016,Diby2017rmp,DEChang2018,Zhanglong2019,Wangquantum2019,Liuquantum2019,Osellamequantum2019,Wangzh20PRL}. Waveguide QED describes interaction phenomena between electromagnetic fields confined to a one-dimensional (1D) waveguide and nearby quantum emitters. In practice, the waveguide can be realized with a number of physical systems such as surface plasmon nanowire \cite{Chang2007nap,AkimovNature2007,Tudela2011prl,Akselrod2014}, diamond waveguide \cite{ClaudonPhoton2010,BabinecNat2010,Clevenson2015,Sipahigil2016},
optical nanofiber \cite{DayanScience2008,Petersen2014,CSayrin2015,PSolano2017,Cheng2017pra,song2017pra,Shinya2019}, photonic crystal \cite{TudelaNAT2015,Burgersa2019,JDHood2016PNAS,JSDouglas2015,Song2018,AGobannatc2014,MArcari2014prl,SPYuapl2014}, and superconducting microwave transmission line \cite{WallraffNature2004,AstafievScience2010,LooSci2013,YLiu2017,Kockum2018,Song2019PRA}. Recently, photon transport in a 1D
waveguide coupled to quantum emitters has been widely studied both in theory \cite{Lzhou2008,HuangPRA2013,Liao2015PRA,Qinwei2016,EWAN2017,Asenjo2017prx,Cheng2018PRA,Litao2018, Xiaky2020} and experiment \cite{GoutaudPRL2015,Vetsch2010prl,NMSun2019}.

Single-photon scattering by multiple emitters has been studied, which gives rise to much richer optical behaviors due to multiple scattering effects. Kien \emph{et al.} \cite{Kien2005} calculated the spontaneous emission from a pair of two-level atoms near a nanofiber, where a substantial radiative exchange between distant atoms was demonstrated. Later, Tsoi and Law \cite{Tsoi2008} studied the interaction between a single photon and a chain of $N$ equally spaced two-level atoms in a 1D waveguide. In contrast to the single-atom case \cite{ShenOL2005,ShenPRL2005}, they found that a photon can be perfectly transmitted near the resonance atomic frequency, and the positions of transmission peaks and their widths are sensitive to the relative position between atoms. Moreover, Chang \emph{et al.}  \cite{Chang2012} showed that an ensemble of periodically arranged two-level atoms with a specific lattice constant can form an effective cavity within the nanofiber. Then, Liao \emph{et al.} \cite{LiaoPRA2016} developed a dynamical theory for calculating photon transport in a 1D waveguide coupled to identical and nonidentical emitters, where the effects of the waveguide
and non-waveguide vacuum modes are included. With a real-space approach,  Zhou \emph{et al.} \cite{ZHouYPRA2017}
investigated the dependence of the single-photon superradiant emission rate on the distance
between atoms in a 1D waveguide. Later, Kornovan \emph{et al.} \cite{Kornovan2019} studied the subradiant collective
states in a periodic 1D array of two-level atoms and showed that long-lived subradiant states
can be obtained with proper system parameters. Recently, Corzo \emph{et al.} \cite{Corzo2019} experimentally
observe a single collective atomic excitation in arrays of individual caesium atoms trapped
along an optical nanofibre. Their work paves the way to herald, store and read out a single
collective atomic excitation in waveguide-QED platforms.

Due to atomic collisions during the loading process, each lattice trap site surrounding a 1D waveguide contains at most a single atom in current experiments \cite{Mitsch2014}. In fact, photon scattering by a waveguide-mediated randomly distributed atomic chain has attracted much attention and been studied both in theory \cite{Mirza2017PRA,Mirza2018josab,Green2019,Kuml2019arxiv,JenPRA2020} and experiment \cite{HLsorensen2016,Corzo2016,SuNJP2019}.
Motivated by these important works, we here focus on the optical properties of an ensemble of two-level atoms coupled to a 1D waveguide.  In our work, we assume that the atoms in our system are randomly trapped in the lattice along the 1D waveguide, which is different from the cases where the atoms are arranged periodically. We calculate the scattering properties of a weak coherent input field through an atomic chain and average over a large sample of atomic distributions.
Provided that the input field is monochromatic, we first study the scattering properties of a two-level atomic chain coupled to a 1D waveguide. The results show that the transport properties are influenced by the lattice constant and the filling factor of the lattice sites. We calculate the optical depth as a function of the lattice constant, concluding that different choices of the lattice constant change the optical depth. Optical depth measures the attenuation of the transmitted field in a material. In quantum memory, the storage efficiency of a medium is determined by the optical depth \cite{Lvovsky2009}. We then focus on the atomic mirror configuration and give the reflection spectra of the incident field with different choices of the filling factors of the lattice sites. Also, we analyze the effect of the inhomogeneous broadening in atomic resonant transition on the scattering spectrum of the input field. Besides, we check the validity of the Markovian assumption and discuss the non-Markovian effect on the scattering spectra when the number of the atoms is large. In a multi-atom system, one mainly focus on the quantum interference between the various fields scattered by atoms. Especially, the interference effect in the photon-photon correlation function is called `quantum beat' (oscillation) \cite{Zheng2013prl}. Studying the persistent quantum beats may reveal the nonlinearity of the atoms in a 1D waveguide. Finally, we calculate the second-order correlation function of the transmitted field with different choices of the filling factors of the lattice sites. We find that quantum beats appear in photon-photon correlation function of the transmitted field. Moreover, when we increase the filling factor of the lattice sites, quantum beat lasts longer. Therefore, the filling factor of the lattice sites provides an efficient way to modify the quantum beats in the second-order correlation function of the transmitted field.

In contrast to a conventional periodic chain of identical emitters, we here
provide a numerical method to deal with the filling imperfection of the atoms in lattice
sites along the 1D waveguide. That is, we take the average values from a large sample of
atomic spatial distributions. In fact, while the lattice sites are periodic, the disorder
caused by the filling imperfections leads to a non-periodic chain of atoms. Due to strong
trap light fields, inhomogeneous broadening of atomic transitions may exist in experiment.
Thus, we here analyze the influence of inhomogeneous broadening on the optical properties
of the incident field. Besides, we also focus on the atomic mirror configuration and give
the reflection spectra of the incident field with imperfect filling of atoms in the lattice
sites. Our numerical results show that, the effects of atomic mirror configuration are
robust to filling imperfections and depend purely on the periodicity of the lattice sites.
That is, the atomic mirror configuration can be also constructed with imperfect filling of
the atoms along a 1D waveguide. By calculating the photon-photon correlation function,
we study the nonlinearity of the atoms mediated by waveguide modes and observe persistent quantum beats.
In a word, our work is a step towards non-periodic atomic chain in waveguide-QED platforms,
and the results give a comprehensive understanding of the linear and
nonlinear optical properties of atom-waveguide system.


\section{MODEL SYSTEM}  \label{MODEL}

\begin{figure}[!ht]
\centering\includegraphics[width=6.5cm]{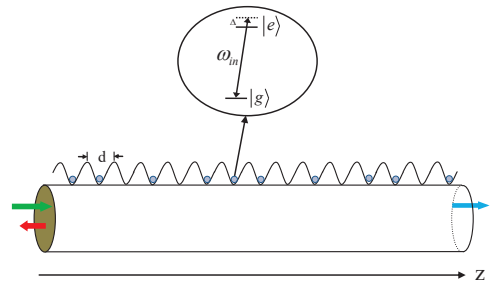}
\caption{(a) Schematic diagram of the propagation of an
input field through a two-level atomic ensemble (grey dots) coupled to a 1D waveguide (the cylinder).
The wavy line denotes the optical lattice external to the waveguide, and the lattice constant is marked
by $d$. A weak coherent field (green arrow) is incident from left to scatter with the atomic ensemble,
which produces output fields including a transmitted part (blue arrow) and a reflected part (red arrow).
In practice, due to the collisional blockade mechanism \cite{Schlosser2002}, either zero
or one atom is trapped in each lattice site \cite{Corzo2016,HLsorensen2016}.} \label{figure1}
\end{figure}

In this section, we consider a system comprising an ensemble of two-level atoms spaced along
a 1D waveguide, as shown in Fig. \ref{figure1}. Each atom has two electronic levels,
i.e., the ground state $|\text{g}\rangle$ and the excited state $|e\rangle$. We assume that the
transition with the resonance frequency $\omega_{a}$ between states $|\text{g}\rangle$ and $|e\rangle$
is coupled to the guided modes of the 1D dielectric waveguide. By generating an optical
lattice external to the waveguide, the atoms can be trapped in fixed positions \cite{Vetsch2010prl,AGoban2015PRL}.
Here, the frequency $\omega_{a}$ is assumed to be away from the waveguide cut-off frequency so that the left-
and right-propagating fields can be treated as completely separate quantum fields \cite{ShenOL2005,Chang2007nap}.
Under rotating wave approximation, the Hamiltonian of the system in real space is given by ($\hbar=1$) \cite{ShenOL2005}

\begin{eqnarray}     \label{eqa1}       
H&\!\!=\!\!&\sum\limits_{j=1}^n\omega_{a}\sigma_{ee}^{j}+iv_{g}\int dz\big[a_{_{L}}^{\dag}(z)\frac{\partial a_{_{L}}(z)}{\partial z}\!-\!a_{_{R}}^{\dag}(z)\frac{\partial a_{_{R}}(z)}{\partial z}\big]\nonumber\\
&&\!\!-\tilde{g}\int dz\sum\limits_{j=1}^n\delta(z-z_{j})\big\{\sigma_{eg}^{j}[a_{_{R}}(z)\!+\!a_{_{L}}(z)]\!+\!\text{H.c.}\big\},
\end{eqnarray}
where $v_{g}$ is the group velocity of the field, $z_{_{j}}$ represents the position of the atom $j$,
and $a_{_{L}}$ ($a_{_{R}}$) denotes the annihilation operator of left (right) propagating field. The
coupling constant $\tilde{g}=\sqrt{2\pi}g$ is assumed to be identical for all modes, where $g$ denotes
the single-atom coupling strength to waveguide modes. The atomic operators $\sigma_{\alpha\beta}^{j}=|\alpha_{j}\rangle\langle\beta_{j}|$,
where $\alpha,\beta=g,e$ are energy eigenstates of the $j$th atom. $n$ is the number of the atoms trapped along the waveguide.

The Heisenberg equation of the motion for the atomic operator is
\begin{eqnarray}     \label{eqa2}       
\dot{\sigma}_{ge}^{j}=-i\omega_{a}\sigma_{ge}^{j}+i\tilde{g}(\sigma_{gg}^{j}-\sigma_{ee}^{j})[a_{_{R}}(z_{j})+a_{_{L}}(z_{j})].
\end{eqnarray}
Likewise, we can also obtain the Heisenberg equations of motions for left and right propagating fields in the waveguide
\begin{eqnarray}     \label{eqa3}       
(\frac{1}{v_{g}}\frac{\partial}{\partial t}-\frac{\partial}{\partial z})a_{_{L}}(z)=\frac{i\tilde{g}}{v_{g}}\sum\limits_{j=1}^n\delta(z-z_{j})\sigma_{ge}^{j},\nonumber\\
(\frac{1}{v_{g}}\frac{\partial}{\partial t}+\frac{\partial}{\partial z})a_{_{R}}(z)=\frac{i\tilde{g}}{v_{g}}\sum\limits_{j=1}^n\delta(z-z_{j})\sigma_{ge}^{j}.
\end{eqnarray}
Then, we transform them to a co-moving frame with coordinates $z'=z$, $t'=t-z/v_{g}$, and get the equation
of motion for $a_{_{R}}$
\begin{eqnarray}     \label{eqa4}       
\frac{\partial}{\partial z'}a_{_{R}}(z')=\frac{i\tilde{g}}{v_{g}}\sum\limits_{j=1}^n\delta(z'-z_{j})\sigma_{ge}^{j}(t').
\end{eqnarray}
Integrating over $z'\in[z-v_{g}t, z]$, we obtain
\begin{eqnarray}     \label{eqa5}       
a_{_{R}}(z,t)\!-\!a_{_{R}}(z\!-\!v_{g}t)=\frac{i\tilde{g}}{v_{g}}\sum\limits_{j=1}^n\int dz'\delta(z'-z_{j})\sigma_{ge}^{j}(t').
\end{eqnarray}
Since the contribution from a time earlier than $z-v_{g}t$ is zero, the lower limit of the integral on
the right hand side of Eq. (\ref{eqa5}) can be extended to $-\infty$. We then get
\begin{eqnarray}     \label{eqa6}       
\begin{split}
a_{_{R}}(z,t)\!=\!a_{_{R,in}}(z\!-\!v_{g}t)\!+\!\frac{i\tilde{g}}{v_{g}}\!\sum\limits_{j}\!\theta(z\!-\!z_{j})\sigma_{ge}^{j}(t\!-\!\frac{z\!-\!z_{j}}{v_{g}}).
\end{split}
\end{eqnarray}
Here $\theta$ represents the Heaviside step function, and $a_{_{R,in}}(z-v_{g}t)$ denotes the input
field which evolves from the initial time to the present without interacting with the atoms. Similarly,
the operator $a_{_{L}}(z,t)$ for the left-moving field is written as
\begin{eqnarray}     \label{eqa7}       
\begin{split}
a_{_{L}}(z,t)\!=\!a_{_{L,in}}(z\!+\!v_{g}t)\!+\!\frac{i\tilde{g}}{v_{g}}\!\sum\limits_{j}\!\theta(z_{j}\!-\!z)\sigma_{ge}^{j}(t\!-\!\frac{z_{j}\!-\!z}{v_{g}}).
\end{split}
\end{eqnarray}
Then, we insert Eq. (\ref{eqa6}) and Eq. (\ref{eqa7}) into Eq. (\ref{eqa2}) and get the evolution of the atomic coherence
\begin{eqnarray}     \label{eqa8}       
\dot{\sigma}_{ge}^{j}=-i\omega_{a}\sigma_{ge}^{j}-\frac{\tilde{g}^{2}}{v_{g}}(\sigma_{gg}^{j}-\sigma_{ee}^{j})\sum\limits_{l}\sigma_{ge}^{l}(t-\frac{|z_{j}-z_{l}|}{v_{g}}).
\end{eqnarray}
By defining new operators $S_{ge}^{j}$ via $\sigma_{ge}^{j}=S_{ge}^{j}e^{-i\omega_{in}t}$,
we transform Eq. (\ref{eqa8}) into a slow-varying frame. Here $\omega_{in}$ denotes the frequency of an external
driving field, which is close to the atomic resonance frequency $\omega_{a}$ with wave vector $k_{a}$. Thus, we find the equation of motion
\begin{eqnarray}     \label{eqa9}       
\dot{S}_{ge}^{j}(t)=\!\!\!&&i\Delta S_{ge}^{j}(t)-\frac{\Gamma_{_{0}}}{2}[S_{gg}^{j}(t)-S_{ee}^{j}(t)]\nonumber\\
&&\times\sum\limits_{l}S_{ge}^{l}(t-\frac{|z_{j}-z_{l}|}{v_{g}})e^{ik_{in}|z_{j}-z_{l}|},
\end{eqnarray}
where $\Delta=\omega_{in}-\omega_{a}$, and $k_{in}=\omega_{in}/v_{g}$. $\Gamma_{_{0}}=4\pi g^2/v_{g}$ denotes
the single-atom spontaneous emission rate into waveguide modes.
In fact, the operator on the right hand side of Eq. (\ref{eqa9}) can be expanded as
\begin{eqnarray}     \label{eqa10}       
S_{ge}^{l}(t-\frac{|z_{j}\!\!-\!\!z_{l}|}{v_{g}})\!\!=\!\!\!\!\!&&S_{ge}^{l}(t)\!\!-\!\!\frac{|z_{j}-z_{l}|}{v_{g}}\dot{S}_{ge}^{l}(t)\nonumber\\
&&+\frac{1}{2}(\frac{|z_{j}-z_{l}|}{v_{g}})^{2}\times{\ddot{S}}_{ge}^{l}(t)+\cdots.
\end{eqnarray}
For small separations, i.e., $|z_{j}-z_{l}|\ll v_{g}$, the system is Markovian \cite{CanevaNJP2015}.
In this case, the time delay for propagation between the atoms can be neglected and so the photon-mediated interactions between atoms occur instantly.
Omitting higher order terms of Eq. (\ref{eqa10}), we get
\begin{eqnarray}     \label{eqa11}       
\dot{S}_{ge}^{j}(t)=\!\!\!\!\!&&-\frac{\Gamma_{_{0}}}{2}[S_{gg}^{j}(t)-S_{ee}^{j}(t)]\nonumber\\
&&\times\sum\limits_{l}S_{ge}^{l}(t)e^{ik_{in}|z_{j}-z_{l}|}+i\Delta S_{ge}^{j}(t).
\end{eqnarray}
From the above equation, we can extract an effective Hamiltonian for our system \cite{Chang2012}
\begin{eqnarray}     \label{eqa12}       
H_{eff}=-\Delta{{\sum\limits_{j=1}^n}} S_{ee}^{j}-i\frac{\Gamma_{_{0}}}{2}{{\sum\limits_{j,k=1}^n}}e^{ik_{a}|z_{_{j}}-z_{_{k}}|}S_{eg}^{j}S_{ge}^{k}.
\end{eqnarray}
Considering the spontaneous emission of the excited state into free space, we can add an imaginary
part $-i\frac{\Gamma_{e}^{'}}{2}$ to the energy of the excited state \cite{Carmichael1993}. Thus,
the atomic chain mediated by the 1D waveguide can be described by a
non-Hermitian effective Hamiltonian \cite{CanevaNJP2015}
\begin{eqnarray}     \label{eqa13}       
H_{1}\!=\!-\!{{\sum\limits_{j=1}^n}}(\Delta\!+\!i\Gamma_{e}^{'}/2)S_{ee}^{j}\!\!-\!i\frac{\Gamma_{_{0}}}{2}\!{{\sum\limits_{j,k=1}^n}}e^{ik_{a}|z_{_{j}}\!-\!z_{_{k}}|}S_{e\text{g}}^{j}S_{\text{g}e}^{k},
\end{eqnarray}
where $\Gamma_{e}^{'}$ denotes the decay rate of the state $|e\rangle$ into free space.

\begin{figure}[!ht]
\centering\includegraphics[width=8.5cm]{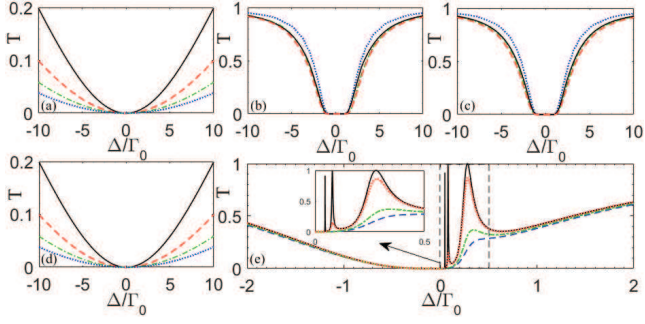}
\caption{ The transmission spectra of the input
field as a function of the frequency detuning $\Delta/\Gamma_{_{0}}$ for (a) $k_{a}d\!=\!0$, (b) $k_{a}d\!=\!1$,
(c) $k_{a}d\!=\!\pi/2$, and (d) $k_{a}d\!=\!\pi$ with filling factors of the lattice sites 0.4 (black solid line),
0.6 (red dashed line), 0.8 (green dashed-dotted line), 1.0 (blue dotted line). (e) The transmission spectra of the input
field for $\Gamma_{e}'=0$ (black solid line), $\Gamma_{e}'=0.01\Gamma_{_{0}}$ (red dotted line), $\Gamma_{e}'=0.1\Gamma_{_{0}}$ (green dashed-dotted line), $\Gamma_{e}'=0.17\Gamma_{_{0}}$ (blue dashed line)
with $N=4$, $p=1$ and $k_{a}d\!=\!0.95\pi$.
Parameters: (a)-(d) $\Gamma_{e}'=0.1\Gamma_{_{0}}$, $N=100$, (a)-(e)
$\mathcal {E}=10^{-4}\sqrt{\frac{\Gamma_{_{0}}}{2v_{g}}}$. } \label{figure2}
\end{figure}

In this work, we mainly study the scattering properties of a weak coherent input field. Then, the
driving part is given by $H_{d}\!=\!\sqrt{\frac{\Gamma_{_{0}}v_{g}}{2}}\mathcal {E}{{\sum\limits_{j=1}^n}}(S_{eg}^{j}e^{ik_{in}z_{_{j}}}+\text{H.c.})$, with $\mathcal {E}$ being the amplitude of the weak input field (Rabi frequency $\sqrt{\frac{\Gamma_{_{0}}v_{g}}{2}}\mathcal {E}$).
Finally, the dynamics of the atomic ensemble is described by the Hamiltonian $H=H_{1}+H_{d}$. Since the
incident field is assumed to be sufficiently weak ($\sqrt{\frac{\Gamma_{_{0}}v_{g}}{2}}\mathcal {E}\!\ll\!\Gamma_{e}^{'}$),
we can neglect quantum jumps \cite{EWAN2017}. Initially, all atoms are prepared in the ground state $|\text{g}\rangle$,
and the weak coherent field is input from the left. Using input-output method \cite{CanevaNJP2015},
we obtain the transmitted ($t$) and reflected ($r$) fields
\begin{eqnarray}     \label{eqa14}       
a_{t}(z) &=& \mathcal {E}e^{ik_{in}z}+i\sqrt{\frac{\Gamma_{_{0}}}{2v_{g}}}{{\sum\limits_{j=1}^n}}S_{ge}^{j}e^{ik_{a}(z-z_{j})},\nonumber\\
a_{r}(z) &=& i\sqrt{\frac{\Gamma_{_{0}}}{2v_{g}}}{{\sum\limits_{j=1}^n}}S_{ge}^{j}e^{-ik_{a}(z-z_{j})}.
\end{eqnarray}
Thus, the transmittance ($T$) and reflection ($R$) of the weak input field are given by
\begin{eqnarray}     \label{eqa15}       
T=\frac{\langle\psi|a_{t}^{\dagger}a_{t}|\psi\rangle}{\mathcal {E}^{2}},\;\;\;\;\;\;\;\;\;
R=\frac{\langle\psi|a_{r}^{\dagger}a_{r}|\psi\rangle}{\mathcal {E}^{2}},
\end{eqnarray}
where $|\psi\rangle$ denotes the steady state of the atomic ensemble.

\section{NUMERICAL RESULTS} \label{RESULTS}

\subsection{Scattering properties of the input field} \label{Transmission}

Here, provided that the incident field is monochromatic, we study the scattering properties
of the weak input field with $N=100$ equally spaced lattice sites along the
1D waveguide. We assume that either zero or one atom is trapped in each lattice site, and all
sites are identical with a filling factor $p$. In other words, $n$ atoms are placed randomly
over $N$ sites with a filling factor $p=n/N$ for each site. In Figs. \ref{figure2}(a)-\ref{figure2}(d), we show transmission
spectra of the incident field as a function of the detuning $\Delta/\Gamma_{_{0}}$ with different values
of lattice constant $d$. For each lattice constant, we present the transmission spectra with four different filling factors, i.e., $p=0.4, 0.6, 0.8, 1.0$. We find that, the transmission spectrum
for $k_{a}d=0$ is identical to that for $k_{a}d=\pi$, and the transmission spectra for $k_{a}d=1$ and
$k_{a}d=\pi/2$ are almost the same. Moreover, after calculating many transmission spectra with different choices
of lattice constant $d$ (not shown), we conclude that different values of $k_{a}d$ do not qualitatively
influence the transmission properties, excluding those very close to $m\pi$ ($m$ is an integer).
Besides, for the case $k_{a}d=m\pi$, when we increase the filling factor, the lineshapes
of the transmission spectra exhibit significant broadening, as shown in Fig. \ref{figure2}(a) and Fig. \ref{figure2}(d). This is because,
the collective decay rates of the atoms into the waveguide modes become enhanced when the filling factor of the lattice sites rises.
Different from the Lorentzian line shape in the transmission spectrum of the single two-level atom case \cite{ShenOL2005}, we find that the transmission for an atomic array is approximately
zero in a window centered at $\Delta=0$ for the cases $k_{a}d=\pi/2$ and $k_{a}d=1.0$ \cite{Mukho2019PRA}, as shown in Figs. \ref{figure2}(b)-\ref{figure2}(c). Besides, as shown in Fig. \ref{figure2}(e), we give the transmission spectra for four choices of the decay rate $\Gamma_{e}'$ with $k_{a}d\!=\!0.95\pi$. The results show that, in the case $\Gamma_{e}'=0$, there are $N-1$ (here $N=4$ in Fig. \ref{figure2}(e)) peaks near the resonance frequency in the transmission spectrum for a linear chain of $N$ atoms, which has been discovered in Ref. \cite{Tsoi2008}. While, when we increase the decay rate $\Gamma_{e}'$ gradually, the peaks will be washed out one by one. For example, for a chain of four atoms, there are two peaks and one peak left in the cases $\Gamma_{e}'=0.01\Gamma_{_{0}}$ and $\Gamma_{e}'=0.1\Gamma_{_{0}}$, respectively. Especially, no peak appears when the decay rate increases to $\Gamma_{e}'=0.17\Gamma_{_{0}}$. Thus, the peaks in the transmission spectra are sensitive to the existence of $\Gamma_{e}'$. To observe these peaks in experiment, the decay rate $\Gamma_{e}'$ needs to be strongly suppressed.

\begin{figure}[!ht]
\centering\includegraphics[width=8.8cm]{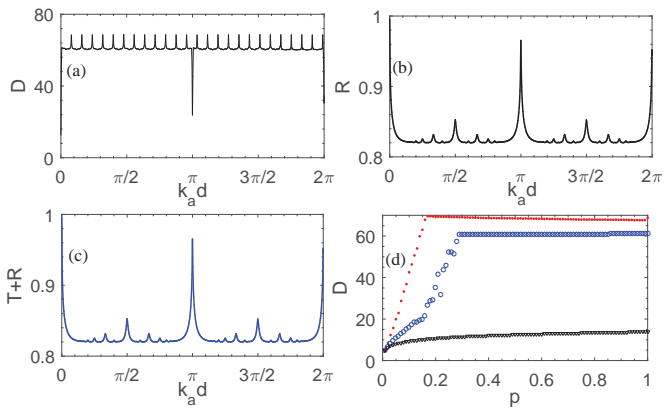}
\caption{(a) The optical depth $D$ versus $k_{a}d$ with a filling factor $p=0.5$. (b) The reflection of the input field as a function of $k_{a}d$ in the resonant case $\Delta=0$ with a filling factor $p=0.5$. (c) Sum of the reflection and transmission of the input field as a function of $k_{a}d$ in the resonant case $\Delta=0$ with a filling factor $p=0.5$.
(d) The optical depth $D$ versus the filling factor $p$ for $k_{a}d=1$ (red dots),  $k_{a}d=\pi/2$ (blue circles) and $k_{a}d=\pi$ (black down-triangles). Parameters: (a)-(d) $\mathcal {E}=10^{-4}\sqrt{\frac{\Gamma_{_{0}}}{2v_{g}}}$, $\Gamma_{e}'=0.1\Gamma_{_{0}}$, $N=100$.
} \label{figure3}
\end{figure}

The opacity of a medium is described by the optical depth $D$, where $T_{\Delta=0}=e^{-D}$.
In Fig. \ref{figure3}(a), we calculate the optical depth as a function of $k_{a}d$
with a filling factor $p=0.5$. For $0< k_{a}d< 2\pi$, the optical depth
is symmetric around $k_{a}d=\pi$, and its minimum is found at $k_{a}d=\pi$.
For $0< k_{a}d< \pi$ or $\pi< k_{a}d< 2\pi$, the optical depth changes with $k_{a}d$ periodically
and gets the maximum at some fixed values of $k_{a}d$.
Obviously, for $k_{a}d=m\pi$, the optical depth is much smaller than that for the
condition $k_{a}d\neq m\pi$. In fact, the case $k_{a}d=m\pi$, with $m$ being an integer,
corresponds to the atomic mirror configuration \cite{Chang2012,Deutsch1995,Mirhosseini2019}.
For clear presentation, in Fig. \ref{figure3}(b), we present the reflection of the input field
in the resonant case $\Delta=0$ as a function of $k_{a}d$ with
the filling factor $p=0.5$. We find that, for $0< k_{a}d< 2\pi$, the reflection spectrum
is symmetric around $k_{a}d=\pi$, and the maximal reflection occurs at $k_{a}d=\pi$.
For $0< k_{a}d< \pi$ ($\pi< k_{a}d< 2\pi$), the reflection spectrum is symmetric around $k_{a}d=0.5\pi$ ($k_{a}d=1.5\pi$).
Besides, the reflection of the input field in the case $k_{a}d=m\pi$ is much larger than those for $k_{a}d\neq m\pi$.
In fact, for the case $k_{a}d=m\pi$, it has been shown that an array of $N$ atoms is
equivalent to an effective `superatom' with $N$ times the coupling strength to the 1D waveguide \cite{ZHouYPRA2017}.
Thus it is possible to use such an atomic ensemble to compensate for the fact that the ratio
$\Gamma_{_{0}}/\Gamma_{e}'$ is not sufficiently large for an individual atom and then the input field is strongly reflected.
Moreover, we give $(T+R)$ of the input field as a function of $k_{a}d$ with $\Delta=0$ and $p=0.5$ in Fig. \ref{figure3}(c). Since the transmission of the input field in the resonant case is nearly zero, Fig. \ref{figure3}(c) is almost identical to Fig. \ref{figure3}(b).
As shown in Fig. \ref{figure3}(d), we also calculate the optical depth $D$ as a function of the filling factor $p$ for three choices of the lattice constant, i.e., $k_{a}d=1, \pi/2, \pi$. The results show that, the variation trends of optical depth in the two cases $k_{a}d=1$ and $k_{a}d=\pi/2$ are similar, i.e., the optical depth first rises nearly linearly with the filling factor and it then becomes saturated when the filling factor increases to a certain value. While for the case $k_{a}d=\pi$, the optical depth changes slowly with the filling factor of the lattice sites. The above results show that, in the case $k_{a}d\neq m\pi$, we can obtain a large optical depth with a modest filling factor. In fact, for a large optical depth, the strong attenuation of the transmitted field is due to the destructive quantum interference between the incident photon field and the re-emitted field by the atomic chain in the forward direction.

\begin{figure}[!ht]
\centering\includegraphics[width=8.8cm]{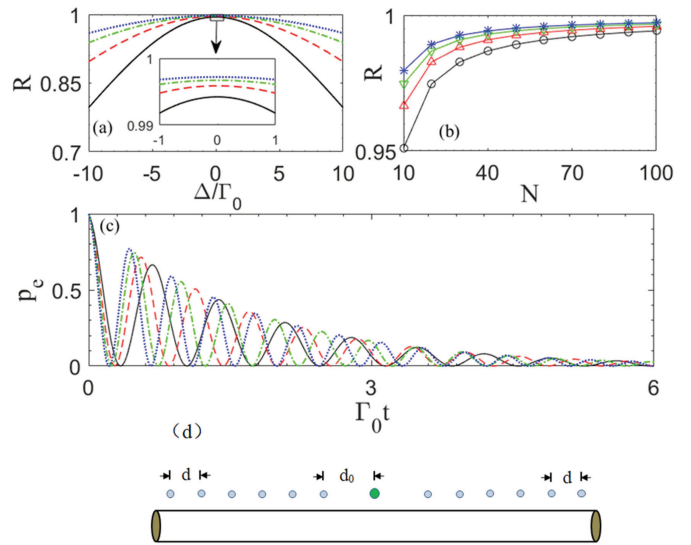}
\caption{(a) Reflection spectra of the input field for the filling factors 0.4 (black solid line), 0.6 (red dashed line), 0.8 (green dashed-dotted line), 1.0 (blue dotted line). (b) The reflection of the input field in the resonant case $\Delta=0$ as a function of the number $N$ of the lattice sites for the filling factors 0.4 (black circles), 0.6 (red up-triangles), 0.8 (green down-triangles), 1.0 (blue asterisks). (c) The population $p_{e}$ of an initially excited atom (green dot in (d) ) inside an atomic cavity (i.e., two sets of atomic Bragg mirrors shown in (d) ) with the filling factors 0.4 (black solid line), 0.6 (red dashed line), 0.8 (green dashed-dotted line), 1.0 (blue dotted line), respectively. (d) An initially excited atom (green dot) inside an atomic cavity (two sets of grey dots) with $k_{a}d_{0}=1.5\pi$ and $k_{a}d=\pi$.  (a)-(c) $\mathcal {E}=10^{-4}\sqrt{\frac{\Gamma_{_{0}}}{2v_{g}}}$, $\Gamma_{e}'=0.1\Gamma_{_{0}}$, $k_{a}d=\pi$. (a) and (c) $N=100$. } \label{figure4}
\end{figure}

\begin{figure}[!ht]
\centering\includegraphics[width=8.6cm]{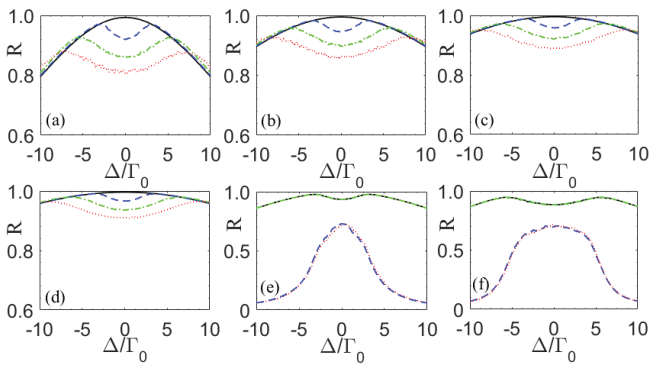}
\caption{Reflection spectra of the incident field as a function of the detuning $\Delta/\Gamma_{_{0}}$ for
$\sigma_{ih}=0$ (black solid line), $\sigma_{ih}=1.0\Gamma_{_{0}}$ (blue dashed line), $\sigma_{ih}=2.0\Gamma_{_{0}}$ (green dashed-dotted line), $\sigma_{ih}=3.0\Gamma_{_{0}}$ (red dotted line) with filling factors (a) $p=0.4$, (b) $p=0.6$, (c) $p=0.8$, (d) $p=1.0$. Reflection spectra of the incident field as a function of the detuning $\Delta/\Gamma_{_{0}}$ for
$\overline{k}_{a}d=0$ (green dashed-dotted line), $\overline{k}_{a}d=1$ (red dotted line), $\overline{k}_{a}d=0.5\pi$ (blue dashed line),  $\overline{k}_{a}d=\pi$ (black solid line)
with (e) $\sigma_{ih}=1.0\Gamma_{_{0}}$ and (f) $\sigma_{ih}=2.0\Gamma_{_{0}}$.
Parameters: (a)-(d) $\overline{k}_{a}d=\pi$,  (a)-(f) $\mathcal {E}=10^{-4}\sqrt{\frac{\Gamma_{_{0}}}{2v_{g}}}$, $\Gamma_{e}'=0.1\Gamma_{_{0}}$, $N=100$, (e)-(f) $p=0.5$. } \label{figure5}
\end{figure}

To proceed, we focus on the atomic mirror configuration, e.g., the case $k_{a}d=\pi$. In Fig. \ref{figure4}(a),
we give reflection spectra of the incident field as a function of the detuning $\Delta/\Gamma_{_{0}}$ with four choices
of filling factor. We find that a high filling factor results in an increase in the
reflection of the input field. In fact, for a resonant input field,
a large number of periodically arranged atoms with a lattice constant $d=m\pi/k_a$
can be regarded as an atomic Bragg mirror \cite{Chang2012,Deutsch1995,Mirhosseini2019}. This effect is not sensitive to the filling imperfections and depends mainly on the periodicity of the lattice sites. Moreover, in the resonant case $\Delta=0$, as shown in Fig. 4(b), the reflection of the atomic Bragg mirror is enhanced by large number $N$ of lattice sites. In fact, for a modest filling factor, when the number of the lattice sites is sufficiently large, the reflection of the atomic Bragg mirror will approach to 100\%. For example, with the filling factor being 0.6, the reflection of the atomic chain in the resonant case is 99.97\% when the number of the lattice sites is 1000 \cite{Corzo2016,HLsorensen2016}. In particular, two sets of such atomic Bragg mirrors can form a cavity for an atom located between them, which is shown in Fig. \ref{figure4}(d). Here, the distance $d_{0}$ between the central atom and the nearest neighbors in the atomic mirrors satisfies the condition $k_{a}d_{0}=1.5\pi$, such that the central atom is
located at the atomic cavity anti-node to maximize the coupling. As shown in Fig. \ref{figure4}(c), we calculate the population
$p_{e}$ of an initially excited atom inside an atomic cavity with four choices of the filling factors. The results
reveal that vacuum Rabi oscillations occur between the excited central atom and the atomic cavity. Moreover, the higher the filling factor of the lattice sites is, the stronger the vacuum Rabi oscillation becomes. This is because the reflection of the atomic cavity rises when we increase the filling factor, which is shown in Fig. \ref{figure4}(a).

\begin{figure}[tpb]    
\centering\includegraphics[width=8.6cm]{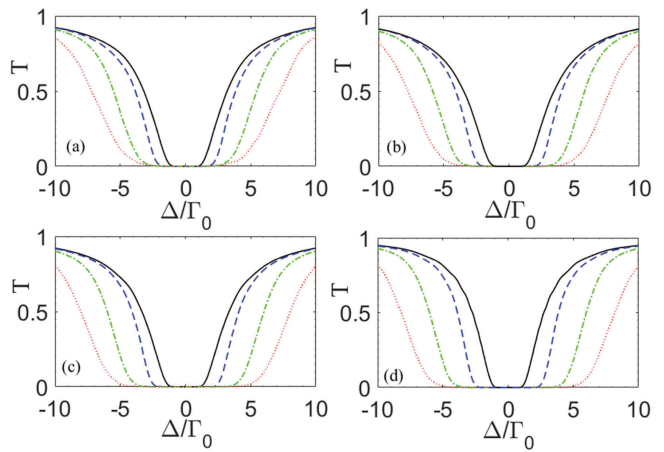}
\caption{Transmission spectra of the incident field as a function of the detuning $\Delta/\Gamma_{_{0}}$ for
$\sigma_{ih}=0$ (black solid line), $\sigma_{ih}=1.0\Gamma_{_{0}}$ (blue dashed line), $\sigma_{ih}=2.0\Gamma_{_{0}}$ (green dashed-dotted line), $\sigma_{ih}=3.0\Gamma_{_{0}}$ (red dotted line) with filling factors (a) $p=0.4$, (b) $p=0.6$, (c) $p=0.8$, (d) $p=1.0$.
Parameters: (a)-(d) $\mathcal {E}=10^{-4}\sqrt{\frac{\Gamma_{_{0}}}{2v_{g}}}$, $\overline{k}_{a}d=\pi/2$, $\Gamma_{e}'=0.1\Gamma_{_{0}}$, $N=100$.  } \label{figure6}
\end{figure}

In the above discussions, we assume that these two-level atoms trapped in the lattice are the same.
While, due to strong trap light fields, inhomogeneous broadening of atomic transitions
exists in practical experiment. Here, we consider the inhomogeneous broadening by assigning a random
Gaussian distributed detuning $\Delta_{ih}$ with a standard deviation $\sigma_{ih}$ to each atom.
The probability density is $\rho_{_{ih}}(\Delta_{ih})=\frac{1}{\sigma_{ih}\sqrt{2\pi}}\exp({-\frac{\Delta_{_{ih}}^{2}}{2\sigma_{ih}^{2}}})$.
As shown in Figs. \ref{figure5}(a)-\ref{figure5}(d), we show the influence of Gaussian inhomogeneous broadening of atomic transitions on the reflection of atomic Bragg mirror, e.g., the case $\overline{k}_{a}d=\pi$. Here, $\overline{k}_{a}$ represents the average of all $k_{a}$. We find that, for a fixed filling factor, when the standard deviation $\sigma_{ih}$ increases, the reflection in a region of the frequency detuning around $\Delta=0$ decreases. That is, the reflection of the atomic cavity shown in Fig. \ref{figure4}(d) is weakened by the inhomogeneous broadening of atomic transitions.
The results also reveal that, for $|\Delta|\gg\Gamma_{0}$, the reflection of the atomic Bragg mirror is almost robust to the exist of the inhomogeneous broadening of the atomic transition. Moreover, as shown in Figs. \ref{figure5}(e)-\ref{figure5}(f), we also calculate the reflection spectra of incident field for two cases $\sigma_{ih}=1.0\Gamma_{_{0}}$ and $\sigma_{ih}=2.0\Gamma_{_{0}}$ with four choices of $\overline{k}_{a}d$, i.e., $\overline{k}_{a}d=0, 1, 0.5\pi, \pi$. The results reveal that, in the two cases $\sigma_{ih}=1.0\Gamma_{_{0}}$ and $\sigma_{ih}=2.0\Gamma_{_{0}}$, different choices
of $\overline{k}_{a}d$ do not qualitatively change the reflection spectra when $\overline{k}_{a}d\neq m\pi$. That is, for a fixed parameter $\sigma_{ih}$, the reflection spectrum is robust to the variation of the lattice constant $d$ in the case $\overline{k}_{a}d\neq m\pi$. Besides, in Fig. \ref{figure6}, we also calculate transmission spectra of the input field for the case $\overline{k}_{a}d=\pi/2$ with four filling factors for $\sigma_{ih}\!=\!0, \Gamma_{_{0}}, 2\Gamma_{_{0}}, 3\Gamma_{_{0}}$.
We find that, for a fixed filling factor, when the standard deviation $\sigma_{ih}$ is changed from
0 to $3.0\Gamma_{_{0}}$, lineshapes of the transmission spectra exhibit significant broadening.

\begin{figure}[tpb]    
\centering\includegraphics[width=8.6cm]{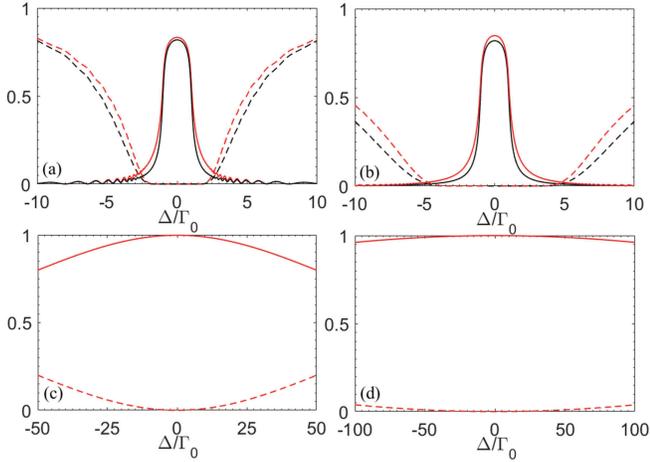}
\caption{Transmission (dashed line) and reflection (solid line) spectra of the incident field as a function of the detuning $\Delta/\Gamma_{_{0}}$ resulting from the effective Hamiltonian with Markovian assumption in this work (black) and the transfer matrix formalism in Ref. \cite{Mukho2019PRA} (red). Note that the red line and black line in (c) or (d) coincide with each other. Parameters: (a) ${k}_{a}d=\pi/2$, $N=100$, (b) ${k}_{a}d=\pi/2$, $N=500$, (c) ${k}_{a}d=\pi$, $N=100$, (d) ${k}_{a}d=\pi$, $N=500$, (a)-(d) $\mathcal {E}=10^{-4}\sqrt{\frac{\Gamma_{_{0}}}{2v_{g}}}$, $\Gamma_{e}'=0.1\Gamma_{_{0}}$, $p=1$, $\sigma_{ih}\!=\!0$.
 } \label{figure7}
\end{figure}

As mentioned in Sec. \ref{MODEL}, when the separations between atoms are small ($|z_{j}-z_{l}|\ll v_{g}$), the system is
Markovian:  the causal propagation time of photons between atoms can be ignored and the interaction between atoms is considered to be instantaneous. While, when we increase of the number of the atoms, this Markovian assumption may be invalid and the atom-waveguide system evolves to be non-Markovian. In fact, the non-Markovian nature of the atomic waveguide system has been widely studied recently \cite{Ballest2013,Tudela2017PRA,Fang2018njp,DincF2019,Sinha2020}. As shown in Fig. \ref{figure7}, we check the validity of the Markovian assumption and discuss the non-Markovian effect in our system. Here, we give the scattering spectra of the incident field for two choices of ${k}_{a}d$ with two methods. one is the input-output theory with the effective Hamiltonian described in Sec. \ref{MODEL}, another is the transfer matrix formalism in Ref. \cite{Mukho2019PRA}. In fact, the deviation between the two results is a sign of non-Markovian behavior. First, we give the transmission and reflection spectra in the case ${k}_{a}d=0.5\pi$ with $N=100$ and $N=500$, as shown in Figs. \ref{figure7}(a)-\ref{figure7}(b). The results reveal that, for the case ${k}_{a}d=0.5\pi$ with $N=100$, although the difference between these two results is macroscopic, it does not qualitatively influence the scattering properties. Then, we increase the number of the atoms to be $N=500$, and observe that the non-Markovian effect on scattering spectra is slightly enhanced. Besides, as shown in Figs. \ref{figure7}(c)-\ref{figure7}(d), we also study the non-Markovian effect in the atomic mirror configuration, e.g., ${k}_{a}d=\pi$.
In this case, the two results in both cases $N=100$ and $N=500$ are identical, which indicates that the non-Markovian effect on the scattering properties disappears in the atomic mirror configuration.
In fact, for the case ${k}_{a}d=m\pi$ with $p=1$, Ref. \cite{Chang2012} has given the analytical results, i.e., $R=\frac{(N\Gamma_{_{0}})^2}{(\Gamma_{e}'+N\Gamma_{_{0}})^2+4\Delta^2}$ and $T=\frac{\Gamma_{e}'^2+4\Delta^2}{(\Gamma_{e}'+N\Gamma_{_{0}})^2+4\Delta^2}$, which are consistent with the results in Ref. \cite{Mukho2019PRA}.

\subsection{Two-photon correlation} \label{correlation}

\begin{figure}[!ht]
\centering\includegraphics[width=8cm]{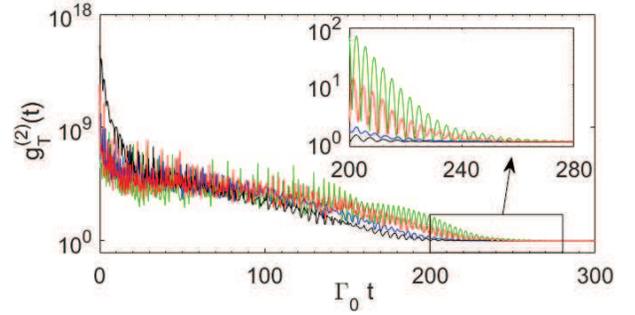}
\caption{The photon-photon correlation function $\text{g}_{_{T}}^{(2)}(t)$ of the transmitted field in the resonant case $\Delta=0$ for the filling factors 0.1 (black line), 0.2 (blue line), 0.3 (red line), 0.4 (green line). Parameters: $\mathcal {E}=10^{-4}\sqrt{\frac{\Gamma_{_{0}}}{2v_{g}}}$, $k_{a}d\!=\!\pi/2$, $N=100$, $\Gamma_{e}'=0.1\Gamma_{_{0}}$, $\sigma_{ih}\!=\!0$. } \label{figure8}
\end{figure}

Correlation between photons is a main feature of nonclassical light, which are characterized
by photon-photon correlation function (second-order correlation function) $g^{2}(t)$ \cite{Loudon2003}.
For a weak coherent state in our system, the photon-photon correlation function $g^{2}$ of the output
field is defined as
\begin{eqnarray}
\text{g}_{\alpha}^{(2)}(\tau)\!\!=\!\!\frac{\langle\psi|a_{_{\alpha}}^{\dagger}(z)e^{iH\tau}a_{_{\alpha}}^{\dagger}(z)a_{_{\alpha}}(z)e^{-iH\tau}a_{_{\alpha}}(z)|\psi\rangle}{|\langle\psi|a_{_{\alpha}}^{\dagger}(z)a_{_{\alpha}}(z)|\psi\rangle|^{2}}.
\end{eqnarray}
Here, $|\psi\rangle$ is the steady-state wave vector and $\alpha\!=\!T, R$.

Now, with a weak input field ($\sqrt{\frac{\Gamma_{_{0}}v_{g}}{2}}\mathcal {E}\!\ll\!\Gamma_{e}^{'}$),
we discuss the photon-photon correlation function of the output field in the resonant case $\Delta=0$.
In Fig. \ref{figure8}, we give the photon correlation function of the transmitted field with
four choices of the filling factor, i.e., $p=0.1,0.2,0.3,0.4$. As shown in Eq. (\ref{eqa14}), the transmitted field arises from
the quantum interference between the incident fields and the forward fields scattered by atoms.
The results show that strong initial bunching appears in the transmitted field in each case, i.e., $\text{g}_{_{T}}^{(2)}(t=0)\gg1$.
When we increase the filling factor, the initial bunching becomes much stronger. Furthermore,
we find persistent quantum beats in the photon-photon correlation function of the transmitted field \cite{Zheng2013prl}. Evidently, higher the filling factor of the lattice sites is, more visible the quantum beat becomes. By comparing these four cases shown in Fig. \ref{figure8}, we find that quantum beat in the photon-photon correlation function $\text{g}_{_{T}}^{(2)}$ lasts
longer when we increase the filling factor. The phenomena mentioned above reveal that
many-body quantum systems significantly modify nonclassical property of light in the waveguide, which is consistent with the results in Refs. \cite{Sahand2018,Sahandarxiv1,Sahandarxiv2}.

\section{Discussion and summary}   \label{discussion}

Experimentally, our model may be realized in the current nanofiber system.
In Refs. \cite{Corzo2016,HLsorensen2016}, arrays of cesium atoms are trapped in the evanescent field of a tapered optical fiber.
For a cesium atom, the ground and excited states are chosen as $|\text{g}\rangle=\{6S_{1/2},F=4\}$ and $|e\rangle\!=\!\{6P_{3/2},F\!=\!5\}$, respectively.
The optical lattice for trapping atoms can be constructed by a pair of horizontally polarized red-detuned counterpropagating beams
(wavelength $\lambda_{trap}=1057$ nm and power $P_{trap}\approx2\times1.3$ mW) and a vertically polarized blue-detuned beam
(wavelength $\lambda_{blue}=780$ nm and power $P_{blue}=14$ mW). In their experiments, cesium atoms are first loaded from a background vapor to a 6-beam magneto-optical trap, and then they are loaded into an optical lattice via sub-Doppler cooling. Because of atomic collisions during the loading process \cite{Schlosser2002}, each trap site in their device hosts at most a single atom, which is consistent with the assumption in the present work. In Ref. \cite{HLsorensen2016}, to avoid saturation, experimentalists adopt an extremely weak probe field with a power of $P_{input}=150$ pW. Finally, with the techniques mentioned above, it is able to trap thousands of atoms in the lattice sites along the 1D waveguide as in Refs. \cite{Corzo2016,HLsorensen2016}.

In conclusion, in this work we study scattering properties of an ensemble of two-level
atoms coupled to a 1D waveguide. Since the precise control of the atomic positions is still challenging
in nanophotonic waveguide system, we assume that the atoms in this work are randomly placed in the
lattice sites along the 1D waveguide. With the effective non-Hermitian Hamiltonian, we calculate the transmission
spectrum of a weak coherent input field, concluding that the optical transport properties are influenced by lattice constant and the filling factor of the lattice sites. We compute the optical depth as a function of the lattice
constant, and the results reveal that the optical depth is reduced when lattice constant is close to $m\pi/k_a$. We
then focus on the atomic mirror configuration and give the reflection spectra of the incident field with different filling factors of the lattice sites. We also quantify the influence of the inhomogeneous broadening in atomic resonant transition on the transmission, and find that the lineshape of the transmission spectrum exhibits significant broadening when the standard deviation $\sigma_{ih}$ becomes larger. Besides, we check the validity of the Markovian assumption adopted in this work and calculate the non-Markovian effect on the scattering properties with a large number of the atoms. Finally, we analyze the role of filling factor played in photon-photon correlation of the transmitted field, and find that initial bunching and quantum beats are sensitive to the filling factor. Since great progress has been made to interface quantum emitters with nanophotonic waveguide \cite{Nayak2018}, the results in this work should be experimentally realizable in the near future.

\section*{ACKNOWLEDGMENTS}


We would like to thank H. R. Wei, Q. Liu, J. Qiu and M. J. Tao for stimulating discussions.
This work is supported by the National Natural Science Foundation of China (11947037, 12004281, 11704214, 11604012);
Program for Innovative Research in University of Tianjin (TD13-5077); Tsinghua University Initiative Scientic
Research Program; Beijing Advanced Innovation Center for Future Chip (ICFC).

\end{CJK*}


\begin{thebibliography}{64}

\bibitem{ShenOL2005} J. T. Shen and S. Fan, Coherent photon transport from spontaneous emission in one-dimensional waveguides, Opt. Lett \textbf{30}, 2001 (2005).



\bibitem{Shen2007PRL}   J. T. Shen and S. Fan, Strongly Correlated Two-Photon Transport in a One-Dimensional Waveguide Coupled to a Two-Level System, Phys. Rev. Lett. \textbf{98}, 153003 (2007).


\bibitem{Zhou2013prl}  L. Zhou, L. P. Yang, Y. Li, and C. P. Sun, Quantum Routing of Single Photons with a Cyclic Three-Level System, Phys. Rev. Lett. \textbf{111}, 103604 (2013).



\bibitem{PLodahl2015rmp}  P. Lodahl, S. Mahmoodian, and S. Stobbe, Interfacing single photons and single quantum dots with photonic nanostructures, Rev. Mod. Phys. \textbf{87}, 347 (2015).


\bibitem{Liao2016}  Z. Liao, X. Zeng, H. Nha, and M. S. Zubairy, Photon transport in a one-dimensional nanophotonic waveguide QED system, Phys. Scr. \textbf{91}, 063004 (2016).


\bibitem{Diby2017rmp}  D. Roy, C. M. Wilson, and O. Firstenberg, Strongly interacting photons in one-dimensional continuum, Rev. Mod. Phys. \textbf{89}, 021001 (2017).



\bibitem{DEChang2018}  D. E. Chang, J. S. Douglas, A. Gonz\'{a}lez-Tudela, C.-L. Hung, and H. J. Kimble, Colloquium: Quantum matter built from nanoscopic lattices of atoms and photons, Rev. Mod. Phys. \textbf{90}, 031002 (2018).


\bibitem{Zhanglong2019} J. Zhang, X. D. Yu, G. L. Long, and Q. K. Xue, Topological dynamical decoupling, Sci. China-Phys. Mech. Astron. \textbf{62}, 120362 (2019).



\bibitem{Wangquantum2019}  M. Wang, R. Wu, J. Lin, J. Zhang, Z. Fang, Z. Chai, and Y. Cheng, Chemo-mechanical polish lithography: A pathway to low
loss large-scale photonic integration on lithium niobate on insulator, Quantum Eng. \textbf{1}, e9 (2019).



\bibitem{Liuquantum2019}  D. E. Liu, Sensing Kondo correlations in a suspended carbon nanotube mechanical resonator with spin-orbit coupling, Quantum Eng. \textbf{1}, e10 (2019).



\bibitem{Osellamequantum2019}   R. Osellame, New effective technique to produce waveguides in lithium niobate on insulator (LNOI), Quantum Eng. \textbf{1}, e11 (2019).



\bibitem{Wangzh20PRL}  Z. H. Wang, T. Jaako, P. Kirton, and P. Rabl, Supercorrelated Radiance in Nonlinear Photonic Waveguides, Phys. Rev. Lett. \textbf{124}, 213601 (2020).


\bibitem{Chang2007nap}  D. E. Chang, A. S. S{\o}rensen, E. A. Demler, and M. D. Lukin, A single-photon transistor using nanoscale surface plasmons, Nat. Phys. \textbf{3}, 807 (2007).



\bibitem{AkimovNature2007}  A. V. Akimov, A. Mukherjee, C. L. Yu, D. E. Chang, A. S. Zibrov, P. R. Hemmer, H. Park, and M. D. Lukin, Generation of single optical plasmons in metallic nanowires coupled to quantum dots, Nature (London) \textbf{450}, 402 (2007).



\bibitem{Tudela2011prl}   A. Gonzalez-Tudela, D. Martin-Cano, E. Moreno, L. Martin-Moreno, C. Tejedor, and F. J. Garcia-Vidal, Entanglement of Two Qubits Mediated by One-Dimensional Plasmonic Waveguides, Phys. Rev. Lett. \textbf{106}, 020501 (2011).


\bibitem{Akselrod2014}   G. M. Akselrod, C. Argyropoulos, T. B. Hoang, C. Cirac\`{\i}, C.
Fang, J. Huang, D. R. Smith, and M. H. Mikkelsen, Probing the mechanisms of large Purcell enhancement in plasmonic nanoantennas, Nature Photonics \textbf{8}, 835 (2014).



\bibitem{ClaudonPhoton2010}   J. Claudon, J. Bleuse, N. S. Malik, M. Bazin, P. Jaffrennou, N. Gregersen, C. Sauvan, P. Lalanne, and J. M. G\'{e}rard, A highly efficient single-photon source based on a quantum dot in a photonic nanowire, Nat. Photon. \textbf{4}, 174 (2010).


\bibitem{BabinecNat2010} T. M. Babinec, J. M. Hausmann, M. Khan, Y. Zhang, J. R. Maze, P. R. Hemmer, and M. Lon\v{c}ar, A diamond nanowire single-photon source, Nat. Nanotechnol. \textbf{5}, 195 (2010).


\bibitem{Clevenson2015}   H. Clevenson, M. E. Trusheim, C. Teale, T. Schr\"{o}der, D. Braje, and D. Englund, Broadband magnetometry and temperature sensing with a light-trapping diamond waveguide, Nat. Phys. \textbf{11}, 393 (2015).


\bibitem{Sipahigil2016}  A. Sipahigil, R. E. Evans, D. D. Sukachev, M. J. Burek, J. Borregaard, M. K. Bhaskar, C. T. Nguyen, J. L. Pacheco,
H. A. Atikian, C. Meuwly, R. M. Camacho, F. Jelezko, E. Bielejec, H. Park, M. Lon\v{c}ar, and M. D. Lukin, An integrated diamond nanophotonics platform for quantum-optical networks, Science \textbf{354}, 847 (2016).


\bibitem{DayanScience2008}   B. Dayan, A. S. Parkins, T. Aoki, E. P. Ostby, K. J. Vahala, and H. J. Kimble, A photon turnstile dynamically regulated by one atom, Science \textbf{319}, 1062 (2008).




\bibitem{Petersen2014}  J. Petersen, J. Volz, and A. Rauschenbeutel, Chiral nanophotonic waveguide interface based on spin-orbit interaction of light, Science \textbf{346}, 67 (2014).



\bibitem{CSayrin2015}  C. Sayrin, C. Clausen, B. Albrecht, P. Schneeweiss, and A. Rauschenbeutel, Storage of fiber-guided light in a nanofiber-trapped ensemble of cold atoms, Optica \textbf{2}, 353 (2015).



\bibitem{PSolano2017}  P. Solano, P. B. Blostein, F. K. Fatemi, L. A. Orozco, and S. L. Rolston, Super-radiance reveals infinite-range dipole interactions through a nanofiber, Nat. Commun. \textbf{8}, 1857 (2017).



\bibitem{song2017pra}   G. Z. Song, E. Munro, W. Nie, F. G. Deng, G. J. Yang, and L. C. Kwek, Photon scattering by an atomic ensemble coupled to a one-dimensional nanophotonic waveguide, Phys. Rev. A \textbf{96}, 043872 (2017).





\bibitem{Cheng2017pra}   M. T. Cheng, J. P. Xu, and G. S. Agarwal, Waveguide transport mediated by strong coupling with atoms, Phys. Rev. A \textbf{95}, 053807 (2017).



\bibitem{Shinya2019}  S. Kato, N. N\'{e}met, K. Senga, S. Mizukami, X. Huang, S. Parkins, and T. Aoki, Observation of dressed states of distant atoms with delocalized photons in coupled-cavities quantum electrodynamics, Nat. Commun. \textbf{10}, 1160 (2019).



\bibitem{AGobannatc2014}  A. Goban, C.-L. Hung, S.-P. Yu, J. D. Hood, J. A. Muniz, J. H. Lee,
M. J. Martin, A. C. McClung, K. S. Choi, D. E. Chang, O. Painter, and H. J. Kimble, Atom-light interactions in photonic crystals, Nat. Commun. \textbf{5}, 3808 (2014).




\bibitem{MArcari2014prl}  M.  Arcari,  I.  S\"{o}llner,  A.  Javadi,  S.  Lindskov  Hansen,  S. Mahmoodian, J. Liu, H. Thyrrestrup, E. H. Lee, J. D. Song, S.  Stobbe,  and  P.  Lodahl, Near-Unity Coupling Efficiency of a Quantum Emitter to a Photonic Crystal Waveguide, Phys.  Rev.  Lett.  \textbf{113},  093603  (2014).




\bibitem{SPYuapl2014}  S.-P. Yu, J. D. Hood, J. A. Muniz, M. J. Martin, R. Norte,
C.-L. Hung, S. M. Meenehan, J. D. Cohen, O. Painter, and H. J.  Kimble, Nanowire photonic crystal waveguides for single-atom trapping and strong light-matter interactions, Appl. Phys. Lett. \textbf{104}, 111103 (2014).



\bibitem{TudelaNAT2015}   A. Gonz\'{a}lez-Tudela, C. L. Hung, D. E. Chang, J. I. Cirac, and H. J. Kimble, Subwavelength vacuum lattices and atom-atom interactions in two-dimensional photonic crystals, Nat. Photonics \textbf{9}, 320 (2015).



\bibitem{JSDouglas2015}  J. S. Douglas, H. Habibian, C. L. Hung, A. V. Gorshkov, H. J. Kimble, and  D. E. Chang, Quantum many-body models with cold atoms coupled to photonic crystals, Nat. Photonics \textbf{9}, 326 (2015).





\bibitem{JDHood2016PNAS}  J. D. Hood, A. Goban, A. Asenjo-Garcia, M. Lu, S.-P. Yu, D. E. Chang, and H. J. Kimble,
Atom-atom interactions around the band edge of a photonic crystal waveguide, Proc. Natl. Acad. Sci. U.S.A. \textbf{113}, 10507 (2016).




\bibitem{Song2018}  G. Z. Song, E. Munro, W. Nie, L. C. Kwek, F. G. Deng, and G. L. Long, Photon transport mediated by an atomic chain trapped along a photonic crystal waveguide, Phys. Rev. A \textbf{98}, 023814 (2018).



\bibitem{Burgersa2019}  A. P. Burgersa, L. S. Penga, J. A. Muniza, A. C. McClunga, M. J. Martina, and H. J. Kimble, Clocked atom delivery to a photonic crystal waveguide, Proc. Natl. Acad. Sci. U.S.A. \textbf{116}, 456 (2019).




\bibitem{WallraffNature2004}   A. Wallraff, D. I. Schuster, A. Blais, L. Frunzio, R. S. Huang, J. Majer, S. Kumar,
S. M. Girvin, and R. J. Schoelkopf, Strong coupling of a single photon to a superconducting qubit using circuit quantum electrodynamics, Nature (London) \textbf{431}, 162 (2004).




\bibitem{AstafievScience2010}  O. Astafiev, A. M. Zagoskin, A. A. Abdumalikov, Y. A. Pashkin,
T. Yamamoto, K. Inomata, Y. Nakamura, and J. S. Tsai, Resonance fluorescence of a single artificial atom, Science \textbf{327}, 840 (2010).



\bibitem{LooSci2013}   A. F. van Loo, A. Fedorov, K. Lalumi\'{e}re, B. C. Sanders, A. Blais, and A. Wallraff, Photon-mediated interactions between distant artificial atoms, Science \textbf{342}, 1494 (2013).



\bibitem{Song2019PRA}  G. Z. Song, L. C. Kwek, F. G. Deng, and G. L. Long, Microwave transmission through an artificial atomic chain coupled to a superconducting photonic crystal, Phys. Rev. A \textbf{99}, 043830 (2019).




\bibitem{YLiu2017}  Y. Liu and A. A. Houck, Quantum electrodynamics near a photonic bandgap, Nat. Phys. \textbf{13}, 48 (2017).



\bibitem{Kockum2018}   A. F. Kockum, G. Johansson, and F. Nori, Decoherence-Free Interaction Between Giant Atoms in Waveguide Quantum Electrodynamics, Phys. Rev. Lett. \textbf{120}, 140404 (2018).





\bibitem{Cheng2018PRA}    D. C. Yang, M. T. Cheng, X. S. Ma, J. P. Xu, C. J. Zhu, and X. S. Huang, Phase-modulated single-photon router, Phys. Rev. A \textbf{98}, 063809 (2018).



\bibitem{EWAN2017}  E. Munro, L. C. Kwek, and D. E. Chang, Optical properties of an atomic ensemble coupled to a band edge of a photonic crystal waveguide, New J. Phys. \textbf{19}, 083018 (2017).




\bibitem{Lzhou2008}  L. Zhou, Z. R. Gong, Y. X. Liu, C. P. Sun, and F. Nori, Controllable Scattering of a Single Photon inside a One-Dimensional Resonator Waveguide, Phys. Rev. Lett. \textbf{101}, 100501 (2008).




\bibitem{Qinwei2016}  W. Qin and F. Nori, Controllable single-photon transport between remote coupled-cavity arrays, Phys. Rev. A \textbf{93}, 032337 (2016).





\bibitem{HuangPRA2013}  J. F. Huang, T. Shi, C. P. Sun, and F. Nori, Controlling single-photon transport in waveguides with finite cross section, Phys. Rev. A \textbf{88},  013836 (2013).



\bibitem{Liao2015PRA}   Z. Liao, X. Zeng, S. Y. Zhu, and M. S. Zubairy, Single-photon transport through an atomic chain coupled to a
one-dimensional nanophotonic waveguide, Phys. Rev. A \textbf{92}, 023806 (2015).


\bibitem{Asenjo2017prx}   A. Asenjo-Garcia, M. Moreno-Cardoner, A. Albrecht, H. J. Kimble, and D. E. Chang, Exponential improvement in
photon storage fidelities using subradiance and selective radiance in atomic arrays, Phys. Rev. X \textbf{7}, 031024 (2017).




\bibitem{Litao2018}   T. Li, A. Miranowicz, X. Hu, K. Y. Xia, and F. Nori, Quantum memory and gates using a $\Lambda$-type quantum emitter
coupled to a chiral waveguide, Phys. Rev. A \textbf{97}, 062318 (2018).



\bibitem{Xiaky2020}   J. S. Tang, Y.Wu, Z. K.Wang, H. Sun, L. Tang, H. Zhang, T. Li, Y. Q. Lu, M. Xiao, and K. Y. Xia, Vacuum-induced
surface-acoustic-wave phonon blockade, Phys. Rev. A \textbf{101}, 053802 (2020).


\bibitem{Vetsch2010prl}  E. Vetsch, D. Reitz, G. Sagu\'{e}, R. Schmidt, S. T. Dawkins, and A. Rauschenbeutel, Optical Interface Created by Laser-Cooled Atoms Trapped in the Evanescent Field Surrounding an Optical Nanofiber, Phys. Rev. Lett. \textbf{104}, 203603 (2010).



\bibitem{GoutaudPRL2015}  B. Gouraud, D. Maxein, A. Nicolas, O. Morin, and J. Laurat, Demonstration of a Memory for Tightly Guided Light in an Optical Nanofiber, Phys. Rev Lett. \textbf{114}, 180503 (2015).



\bibitem{NMSun2019}  N. M. Sundaresan, R. Lundgren, G. Zhu, A. V. Gorshkov, and A. A. Houck, Interacting Qubit-Photon Bound States with Superconducting Circuits, Phys. Rev. X \textbf{9}, 011021 (2019).



\bibitem{Kien2005}  F. Le Kien, S. Dutta Gupta, K. P. Nayak, and K. Hakuta, Nanofiber-mediated radiative transfer between two distant atoms,  Phys. Rev. A \textbf{72}, 063815 (2005).



\bibitem{Tsoi2008}   T. S. Tsoi and C. K. Law, Quantum interference effects of a single photon interacting with an atomic chain inside a one-dimensional waveguide, Phys. Rev. A \textbf{78}, 063832 (2008).


\bibitem{ShenPRL2005}  J. T. Shen and S. Fan, Coherent Single Photon Transport in a One-Dimensional Waveguide Coupled with Superconducting Quantum Bits, Phys. Rev. Lett. \textbf{95}, 213001 (2005).




\bibitem{Chang2012}   D. E. Chang, L. Jiang, A. V. Gorshkov, and H. J. Kimble, Cavity QED with atomic mirrors, New J. Phys. \textbf{14}, 063003 (2012).




\bibitem{LiaoPRA2016}    Z. Liao, H. Nha, and M. S. Zubairy, Dynamical theory of single photon transport in a one-dimensional waveguide coupled to identical and nonidentical emitters, Phys. Rev. A \textbf{94}, 053842 (2016).



\bibitem{ZHouYPRA2017}   Y. Zhou, Z. H. Chen, and J. T. Shen, Single-photon superradiant emission rate scaling for atoms trapped in a photonic waveguide, Phys. Rev. A \textbf{95}, 043832 (2017).





\bibitem{Kornovan2019}   D. F. Kornovan, N. V. Corzo, J. Laurat, and A. S. Sheremet, Extremely subradiant states in a periodic one-dimensional atomic array, Phys. Rev. A \textbf{100}, 063832 (2019).




\bibitem{Corzo2019}    N. V. Corzo, J. Raskop, A. Chandra, A. S. Sheremet, B. Gouraud, and J. Laurat, Waveguide-coupled single collective excitation of atomic arrays, Nature \textbf{566}, 359 (2019).



\bibitem{Mitsch2014}  R. Mitsch, C. Sayrin, B. Albrecht, P. Schneeweiss, and A. Rauschenbeutel, Quantum state-controlled directional spontaneous emission of photons into a nanophotonic waveguide, Nat. Commun. \textbf{5}, 5713 (2014).




\bibitem{Mirza2017PRA}    I. M. Mirza, J. G. Hoskins, and J. C. Schotland, Chirality, band structure, and localization in waveguide quantum
electrodynamics, Phys. Rev. A \textbf{96}, 053804 (2017).



\bibitem{Mirza2018josab}    I. M. Mirza and J. C. Schotland, Influence of disorder on electromagnetically induced transparency in chiral
waveguide quantum electrodynamics, J. Opt. Soc. Am. B \textbf{35}, 1149-1158 (2018).



\bibitem{Green2019}    Y. S. Greenberg and A. G. Moiseev, Influence of impurity on the rate of single photon superradiance and photon
transport in disordered N qubit chain, Phys. E (Amsterdam, Neth.) \textbf{108}, 300 (2019).






\bibitem{Kuml2019arxiv}    J. Kumlin, K. Kleinbeck, N. Stiesdal, H. Busche, S. Hofferberth, and H. S. Buchler, Nonexponential decay of a
collective excitation in an atomic ensemble coupled to a one-dimensional waveguide, Phys. Rev. A \textbf{102}, 063703 (2020).



\bibitem{JenPRA2020}    H. H. Jen, Disorder-assisted excitation localization in chirally coupled quantum emitters, Phys. Rev. A \textbf{102}, 043525 (2020).



\bibitem{Corzo2016}  N. V. Corzo, B. Gouraud, A. Chandra, A. Goban, A. S. Sheremet, D. V. Kupriyanov, and J. Laurat, Large Bragg Reflection from One-Dimensional Chains of Trapped Atoms Near a Nanoscale Waveguide, Phys. Rev. Lett. \textbf{117}, 133603 (2016).




\bibitem{HLsorensen2016}  H. L. S{\o}rensen, J. B. B\'{e}guin, K. W. Kluge, I. Iakoupov, A. S. S{\o}rensen, J. H. M\"{u}ller, E. S. Polzik, and J. Appel, Coherent Backscattering of Light Off One-Dimensional Atomic Strings, Phys. Rev. Lett. \textbf{117}, 133604 (2016).



\bibitem{SuNJP2019}   D. Su, R. Liu, Z. Ji, X. Qi, Z. Song, Y. Zhao, L. Xiao, and S. Jia, Observation of ladder-type electromagnetically
induced transparency with atomic optical lattices near a nanofiber, New J. Phys. \textbf{21}, 043053 (2019).


\bibitem{Lvovsky2009}     A. I. Lvovsky, B. C. Sanders, and W. Tittel, Optical quantum memory, Nat. Photonics \textbf{3}, 706 (2009).


\bibitem{Zheng2013prl}  H. Zheng and H. U. Baranger, Persistent Quantum Beats and Long-Distance Entanglement from Waveguide-Mediated Interactions, Phys. Rev. Lett. \textbf{110}, 113601 (2013).



\bibitem{Schlosser2002}  N. Schlosser, G. Reymond, and P. Grangier, Collisional Blockade in Microscopic Optical Dipole Traps, Phys. Rev. Lett. \textbf{89}, 023005 (2002).




\bibitem{AGoban2015PRL}  A. Goban, C.-L. Hung, J. D. Hood, S.-P. Yu, J. A. Muniz, O. Painter, and H. J. Kimble, Superradiance for Atoms Trapped along a Photonic Crystal Waveguide, Phys. Rev. Lett. \textbf{115}, 063601 (2015).



\bibitem{CanevaNJP2015}   T. Caneva, M. T. Manzoni, T. Shi, J. S. Douglas, J. I. Cirac, and D. E. Chang, Quantum dynamics of propagating photons with strong interactions: a generalized input-output formalism, New J. Phys. \textbf{17}, 113001 (2015).



\bibitem{Carmichael1993}   H. J. Carmichael, \emph{An Open Systems Approach to Quantum Optics } (Springer, Berlin, 1993).


\bibitem{Mukho2019PRA}   D. Mukhopadhyay and G. S. Agarwal, Multiple Fano interferences due to waveguide-mediated phase coupling between atoms, Phys. Rev. A \textbf{100}, 013812 (2019).


\bibitem{Deutsch1995}  I. H. Deutsch, R. J. C. Spreeuw, S. L. Rolston, and W. D. Phillips, Photonic band gaps in optical lattices, Phys. Rev. A \textbf{52}, 1394 (1995).



\bibitem{Mirhosseini2019}  M. Mirhosseini, E. Kim, X. Zhang, A. Sipahigil, P. B. Dieterle, A. J. Keller, A. Asenjo-Garcia, D. E. Chang, and
O. Painter, Cavity quantum electrodynamics with atom-like mirrors, Nature (London) \textbf{569}, 692 (2019).



\bibitem{Ballest2013}    C. Gonzalez-Ballestero, F. J. Garc\'{i}a-Vidal, and E. Moreno, Non-Markovian effects in waveguide-mediated
entanglement, New J. Phys. \textbf{15}, 073015 (2013).



\bibitem{Tudela2017PRA}     A. Gonz\'{a}lez-Tudela and J. I. Cirac, Markovian and non-Markovian dynamics of quantum emitters coupled to
two-dimensional structured reservoirs, Phys. Rev. A \textbf{96}, 043811 (2017).



\bibitem{Fang2018njp}     Y. L. L. Fang, F. Ciccarello, and H. U. Baranger, Non-Markovian dynamics of a qubit due to single-photon scattering
in a waveguide, New J. Phys. \textbf{20}, 043035 (2018).



\bibitem{DincF2019}     F. Dinc, I. Ercan, and A. M. Branczyk, Exact Markovian and non-Markovian time dynamics in waveguide QED:
collective interactions, bound states in continuum, superradiance and subradiance, Quantum \textbf{3}, 213 (2019).



\bibitem{Sinha2020}     K. Sinha, P. Meystre, E. A. Goldschmidt, F. K. Fatemi, S. L. Rolston, and P. Solano, Non-Markovian collective
emission from macroscopically separated emitters, Phys. Rev. Lett. \textbf{124}, 043603 (2020).


\bibitem{Loudon2003}   R. Loudon, The Quantum Theory of Light, 3rd ed. (Oxford University Press, New York, 2003).


\bibitem{Sahand2018}  S. Mahmoodian, M. \u{C}epulkovskis, S. Das, P. Lodahl, K. Hammerer, and A. S. S{\o}rensen, Strongly Correlated Photon
Transport in Waveguide Electrodynamics with Weakly Coupled Emitters, Phys. Rev. Lett. \textbf{121}, 143601 (2018).



\bibitem{Sahandarxiv1} S. Das, L. Zhai, M. \u{C}epulskovskis, A. Javadi, S. Mahmoodian, P. Lodahl, A. S. S{\o}rensen, A wave-function ansatz
method for calculating field correlations and its application to the study of spectral filtering and quantum dynamics of multi-emitter systems, arXiv:1912.08303.



\bibitem{Sahandarxiv2} A. S. Prasad, J. Hinney, S. Mahmoodian, K. Hammerer, S. Rind, P. Schneeweiss, A. S. S{\o}rensen,
J. Volz, Arno Rauschenbeutel, Correlating photons using the collective nonlinear response of atoms weakly coupled to an optical mode, arXiv:1911.09701.



\bibitem{Nayak2018}   K. P. Nayak, M. Sadgrove, R. Yalla, F. Le Kien, and K. Hakuta, Nanofiber quantum photonics, J. Opt. \textbf{20}, 073001 (2018).





\end{thebibliography}
\end{document}